\documentclass{symmetry12}

\begin{document}

\setcounter{page}{71}

\FirstPageHeading{Cardoso-Bihlo}

\ShortArticleName{Differential invariants for the KdV equation}
\ArticleName{Differential Invariants\\ for the Korteweg--de Vries Equation}

\Author{Elsa Maria DOS SANTOS CARDOSO-BIHLO}
\AuthorNameForHeading{E.M. Dos Santos Cardoso-Bihlo}
\AuthorNameForContents{Dos Santos Cardoso-Bihlo E.M.}
\ArticleNameForContents{Differential invariants for the Korteweg--de Vries\\ equation}

\Address{Faculty of Mathematics, University of Vienna, Nordbergstra{\ss}e 15,\\ A-1090 Vienna, Austria}
\Email{elsa.cardoso@univie.ac.at}

\Abstract{Differential invariants for the maximal Lie invariance group of the Korteweg--de Vries equation are computed using the moving frame method and compared with existing results. Closed forms of differential invariants of any order are presented for two sets of normalization conditions. Minimal bases of differential invariants associated with the chosen normalization conditions are given.}

\section{Introduction}

Invariants and differential invariants are important objects associated with transformation groups. They play a role for finding invariant, partially invariant and differentially invariant solutions~\cite{CardosoBihlo:golo04Ay,CardosoBihlo:olve86Ay,CardosoBihlo:ovsi82Ay}, in computer vision~\cite{CardosoBihlo:olve04Ay}, for the construction of invariant discretization schemes~\cite{CardosoBihlo:bihl12By,CardosoBihlo:bihl12Cy,CardosoBihlo:doro11Ay,CardosoBihlo:kim08Ay,CardosoBihlo:levi06Ay,CardosoBihlo:olve01Ay,CardosoBihlo:rebe11Ay} and in the study of invariant parameterization schemes~\cite{CardosoBihlo:bihl12Ey,CardosoBihlo:bihl11Fy,CardosoBihlo:popo10Cy}.

There are two main ways to construct differential invariants for Lie group actions. The notation we use follows the book~\cite{CardosoBihlo:olve86Ay} and the papers \cite{CardosoBihlo:cheh08Ay,CardosoBihlo:fels98Ay,CardosoBihlo:olve01Ay,CardosoBihlo:olve04Ay,CardosoBihlo:olve10Ay,CardosoBihlo:olve08Ay}. Let $G$ be a (pseudo)group of transformations acting on the space of variables $(x,u)$, where $x=(x^1,\dots,x^p)$ is the tuple of independent variables and $u=(u^1,\dots,u^q)$ is the tuple of dependent variables. Let $\mathfrak g$ be the Lie algebra of vector fields that is associated with $G$.

The first way for the computation of differential invariants uses the infinitesimal method~\cite{CardosoBihlo:chup04Ay,CardosoBihlo:golo04Ay,CardosoBihlo:olve86Ay,CardosoBihlo:ovsi82Ay}. The criterion for a function $I$ defined on a subset of the corresponding $n$th-order jet space to be a differential invariant of the maximal Lie invariance group $G$ is that the condition
\begin{equation}\label{CardosoBihlo:InfinitesimalInvarianceCriterion}
 \mathrm{pr}^{(n)}\mathbf v(I)=0,
\end{equation}
holds for any vector field $\mathbf v\in \mathfrak g$. In equation~\eqref{CardosoBihlo:InfinitesimalInvarianceCriterion}, the vector field $\mathbf v$ is of the form $\mathbf v=\xi^i(x,u)\partial_{x^i}+\phi_\alpha(x,u)\partial_{u^\alpha}$  (the summation over double indices is applied), and $\mathrm{pr}^{(n)}\mathbf v$ denotes the standard $n$th prolongation of $\mathbf v$. In the framework of the infinitesimal method, the differential invariants $I$ are computed by solving the system of quasilinear first-order partial differential equations of the form~\eqref{CardosoBihlo:InfinitesimalInvarianceCriterion}, where the vector field~$\mathbf v$ runs through a generating set of~$\mathfrak g$.

The second possibility for computing differential invariants uses moving frames \cite{CardosoBihlo:cheh08Ay,CardosoBihlo:fels98Ay,CardosoBihlo:fels99Ay}. The main advantage of the moving frame method is that it avoids the integration of differential equations, which is necessary in the infinitesimal approach. At the same time, using moving frames allows one to invoke the powerful recurrence relations, which can be helpful in studying the structure of the algebra of differential invariants.

In this paper, we study differential invariants for the maximal Lie invariance group of the Korteweg--de Vries (KdV) equation. This problem was already considered in~\cite{CardosoBihlo:cheh08Ay,CardosoBihlo:olve10Ay} and in~\cite{CardosoBihlo:chup04Ay} within the framework of the moving frame and infinitesimal approaches, respectively. Thus, on one hand it is instructive to compare and review the results available in the literature. On the other hand, we extend these results in the present paper. In particular, we explicitly present functional bases of differential invariants of arbitrary order for the aforementioned group.

The further organization of the paper is the following. In Section~\ref{CardosoBihlo:SymmetriesKdV} we restate the maximal Lie invariance group of the KdV equation. Section~\ref{CardosoBihlo:MovingFrameKdV} collects some results related to a moving frame for the maximal Lie invariance group of the KdV equation as presented in~\cite{CardosoBihlo:cheh08Ay}. We also introduce an alternative moving frame in this section. Section~\ref{CardosoBihlo:DifferentialInvariantsKdV} contains our main results, which are a complete list of functionally independent differential invariants for the maximal Lie invariance group of KdV equation of \textit{any} order as well as the description of a basis of differential invariants for the new normalization introduced in Section~\ref{CardosoBihlo:MovingFrameKdV}. Section~\ref{CardosoBihlo:ConclusionKdV} contains some remarks related to the results of the paper.

\section{Lie symmetries of the KdV equation}\label{CardosoBihlo:SymmetriesKdV}

The KdV equation is undoubtedly one of the most important partial differential equations in mathematical physics. It describes the motion of long shallow-water waves in a channel. Here we will use it in the following dimensionless form:
\begin{equation}\label{CardosoBihlo:KdVequation}
 u_t+uu_x+u_{xxx}=0.
\end{equation}
The KdV equation is completely integrable using inverse scattering~\cite{CardosoBihlo:gard67Ay}.
The coefficients of each vector field~$Q=\tau(t,x,u)\partial_t+\xi(t,x,u)\partial_x+\eta(t,x,u)\partial_u$
generating a one-parameter Lie symmetry group of the KdV equation
satisfy the system of determining equations
\begin{equation}\label{CardosoBihlo:DetEqs}
\tau_x=\tau_u=\xi_u=\eta_t=\eta_x=0,\quad \eta=\xi_t-\tfrac23u\tau_t,\quad \eta_u=-\tfrac23\tau_t=-2\xi_x
\end{equation}
with the general solution
\[
\tau=3c_4t+c_1,\quad \xi=c_4x+c_3t+c_2,\quad \eta=-2c_4u+c_3,
\]
where $c_1, \dots, c_4$ are arbitrary constants.
Hence the maximal Lie invariance algebra~$\mathfrak g$ of~\eqref{CardosoBihlo:KdVequation} is spanned by the four vector fields
\begin{equation}\label{CardosoBihlo:KdVequationAlgebra}
 \partial_t,\quad \partial_x,\quad t\partial_x+\partial_u,\quad 3t\partial_t+x\partial_x-2u\partial_u.
\end{equation}
Associated with these basis elements are the one-parameter symmetry groups of (i)~time translations, (ii)~space translations, (iii)~Galilean boosts and (iv)~\mbox{scalings}. The most general Lie symmetry transformation of the KdV equation can be constructed using these elementary one-parameter groups:
\begin{equation}\label{CardosoBihlo:TransformationFormulasKdV}
T=e^{3\varepsilon_4}(t+\varepsilon_2),\quad
X=e^{\varepsilon_4}(x+\varepsilon_2+\varepsilon_1\varepsilon_3+\varepsilon_3t),\quad
U=e^{-2\varepsilon_4}(u+\varepsilon_3),
\end{equation}
where $\varepsilon_1,\dots,\varepsilon_4\in\mathbb{R}$ are continuous group parameters. The KdV equation also admits a discrete point symmetry, given by simultaneous changes of the signs of the variables~$t$ and~$x$.

The prolongation of the general element~$Q$ of the algebra~$\mathfrak g$ has
\[
\eta^\alpha=-(3\alpha_1+\alpha_2+2)c_4u_\alpha-\alpha_1c_3u_{\alpha_1-1,\alpha_2+1},
\]
as the coefficient of~$\partial_{u_\alpha}$,
where $\alpha=(\alpha_1,\alpha_2)$ is a multiindex, $\alpha_1,\alpha_2\in\mathbb N\cup\{0\}$, and $u_{\alpha}=\partial^{\alpha_1+\alpha_2}u/\partial t^{\alpha_1}\partial x^{\alpha_2}$ as usual.

Using the chain rule, from the above transformation formula~\eqref{CardosoBihlo:TransformationFormulasKdV} one obtains the expressions for the transformed derivative operators,
\[
 \mathrm D_T=e^{-3\varepsilon_4}(\mathrm D_t-\varepsilon_3\mathrm D_x),\quad \mathrm D_X=e^{-\varepsilon_4}\mathrm D_x.
\]
In~\cite{CardosoBihlo:cheh08Ay} these operators were used for listing some of the lower order transformed partial derivatives of~$u$. However, in order to obtain a closed formula for a functional basis of differential invariants of \emph{arbitrary} order for the KdV equation, it is useful to attempt to derive a closed-form expression for the transformed derivatives of~$u$. Such an expression is
\begin{align}\label{CardosoBihlo:TranformedDerivativesKdV}
\begin{split}
U_{\alpha}&=e^{-(3\alpha_1+\alpha_2+2)\varepsilon_4}(\mathrm D_t-\varepsilon_3\mathrm D_x)^{\alpha_1}\mathrm D_x^{\alpha_2}u {}\\ &{} = e^{-(3\alpha_1+\alpha_2+2)\varepsilon_4}\sum_{k=0}^{\alpha_1}(-\varepsilon_3)^k\left(\alpha_1\atop k\right)u_{\alpha_1-k,\alpha_2+k}.
\end{split}
\end{align}
 In particular, the expressions for $U_T$ and $U_X$ are
\[
 U_T=e^{-5\varepsilon_4}(u_t-\varepsilon_3u_x),\quad U_X=e^{-3\varepsilon_4}u_x.
\]

\section{A moving frame for the KdV equation}\label{CardosoBihlo:MovingFrameKdV}

As the maximal Lie invariance group of the KdV equation is finite-dimensional, we only review the construction of moving frames for finite-dimensional group actions here. Details on the moving frame construction for Lie pseudogroups can be found, e.g., in~\cite{CardosoBihlo:cheh08Ay,CardosoBihlo:olve08Ay}.

\begin{definition}
 Let there be given a Lie group $G$ acting on a manifold $M$. A \emph{right moving frame} is a mapping $\rho\colon M\to G$ that satisfies the property $\rho(g\cdot z)=\rho(z)g^{-1}$ for any $g\in G$ and $z\in M$.
\end{definition}

The theorem on moving frames, see e.g.~\cite{CardosoBihlo:fels99Ay,CardosoBihlo:olve04Ay,CardosoBihlo:olve10Ay}, guarantees the existence of a moving frame in the neighborhood of a point $z\in M$ if and only if $G$ acts freely and regularly near $z$. Moving frames are constructed using a procedure called normalization, which is based on the selection of a submanifold (the cross-section) that intersects the group orbits only once and transversally.

There exist infinitely many possibilities to construct a moving frame. The single moving frames differ in the choice of the respective cross-sections. The moving frame constructed in~\cite{CardosoBihlo:cheh08Ay} rests on the normalization conditions
\begin{equation}\label{CardosoBihlo:NormalizationsKdVequation}
 T=0,\quad X=0,\quad U=0,\quad U_T=1,
\end{equation}
i.e.,\ it is defined on the first jet space $J^1$. It is necessary to construct the moving frame on the first jet space, as the maximal Lie invariance group of the KdV equation does not act freely on the space $M$, spanned by $t$, $x$ and $u$. The action of $G$ first becomes free when prolonged to $J^1$, which is then the proper space to construct the moving frame $\rho^{(1)}\colon J^1\to G$ on. Solving the above algebraic system~\eqref{CardosoBihlo:NormalizationsKdVequation} for the group parameters $\varepsilon_1,\dots,\varepsilon_4$ yields the moving frame $\rho^{(1)}$
\begin{equation}\label{CardosoBihlo:MovingFrameKdVequation}
 \varepsilon_1=-t,\quad \varepsilon_2=-x,\quad \varepsilon_3=-u,\quad \varepsilon_4=\frac15\ln(u_t+uu_x),
\end{equation}
which is well defined provided that $u_t+uu_x>0$. This moving frame becomes singular when $u_t+uu_x=0$. The latter condition is equivalent, on the manifold of the KdV equation, to the condition that $u_{xxx}=0$ and implies, together with the KdV equation, that $u_{xx}=0$.

Another possible normalization, leading to an alternative moving frame, is the following:
\[
 T=0,\quad X=0,\quad U=0,\quad U_X=1.
\]
Solving the normalization conditions gives the associated moving frame
\begin{equation}\label{CardosoBihlo:MovingFrameKdVequation2}
 \varepsilon_1=-t,\quad \varepsilon_2=-x,\quad \varepsilon_3=-u,\quad \varepsilon_4=\frac13\ln u_x,
\end{equation}
which is well defined provided that $u_x>0$.

Note that for $u_t+uu_x<0$ (resp. $u_x<0$) one can replace the condition $U_T=1$ by $U_T=-1$ (resp. $U_X=1$ by $U_X=-1$).

\section{Differential invariants for the KdV equation}\label{CardosoBihlo:DifferentialInvariantsKdV}

The above moving frames can now be used to construct differential invariants using the \emph{method of invariantization}~\cite{CardosoBihlo:cheh08Ay,CardosoBihlo:olve04Ay,CardosoBihlo:olve10Ay}.

\begin{definition}
 The \emph{invariantization} of a function $f\colon M\to\mathbb{R}$ is the function defined by
 \[
 \iota(f)=f(\rho(z)\cdot z).
 \]
\end{definition}

We first construct the set of all functionally independent differential invariants for the maximal Lie invariance group of the KdV equation using the moving frame~\eqref{CardosoBihlo:MovingFrameKdVequation}. An exhaustive list of differential invariants of any order was not given in~\cite{CardosoBihlo:cheh08Ay}. Such a list is obtained by plugging the moving frame~\eqref{CardosoBihlo:MovingFrameKdVequation} into the transformed derivatives~\eqref{CardosoBihlo:TranformedDerivativesKdV}. This yields
\begin{equation}\label{CardosoBihlo:DifferentialInvariantsKdVFirstNormalization}
 I_\alpha=\iota(U_\alpha)=(u_t+uu_x)^{-(3\alpha_1+\alpha_2+2)/5}\sum_{k=0}^{\alpha_1} \left(\alpha_1\atop k\right)u^ku_{\alpha_1-k,\alpha_2+k},
\end{equation}
where $\alpha_1>1$ or $\alpha_2>0$. Invariantizing $t$, $x$, $u$ and $u_t$, one recovers the normalization conditions~\eqref{CardosoBihlo:NormalizationsKdVequation} and the associated differential invariants are dubbed \emph{phantom invariants}.
The corresponding invariantized form of the KdV equation is $1+I_{03}=0$.

Using the alternative moving frame~\eqref{CardosoBihlo:MovingFrameKdVequation2}, invariantization of~\eqref{CardosoBihlo:TranformedDerivativesKdV} leads to the following set of functionally independent differential invariants of the maximal Lie invariance group of the KdV equation,
\begin{equation}\label{CardosoBihlo:DifferentialInvariantsKdVSecondNormalization}
 I_{\alpha}=\iota(U_\alpha)=u_x^{-(3\alpha_1+\alpha_2+2)/3}\sum_{k=0}^{\alpha_1}\left(\alpha_1\atop k\right)u^ku_{\alpha_1-k,\alpha_2+k},
\end{equation}
where $\alpha_1>0$ or $\alpha_2>1$, and $H^1=\iota(t)=0$, $H^2=\iota(x)=0$, $I_{00}=\iota(u)=0$ and $I_{01}=\iota(u_x)=1$ exhaust the set phantom invariants for this moving frame.
Then the invariantization of the KdV equation yields the invariant form $I_{10}+I_{03}=0$. The advantage of the form~\eqref{CardosoBihlo:DifferentialInvariantsKdVSecondNormalization} of differential invariants compared to the form~\eqref{CardosoBihlo:DifferentialInvariantsKdVFirstNormalization}, which follows from the normalization~\eqref{CardosoBihlo:NormalizationsKdVequation} chosen in~\cite{CardosoBihlo:cheh08Ay}, is that these invariants are singular only on the subset $u_x=0$, which is contained in the subset $u_{xx}=0$ on which the invariants~\eqref{CardosoBihlo:DifferentialInvariantsKdVFirstNormalization} are singular (again, when restrict to the KdV equation).

In principle, by computing the form of differential invariants of any order we have already solved the problem to exhaustively describe all the differential invariants for the maximal Lie invariance group of the KdV equation. On the other hand, it is instructive to study the structure of the algebra of differential invariants in some more detail.

In particular, an interesting open problem in the theory of differential invariants is to find minimal generating set of differential invariants in an algorithmic way. This is the set of differential invariants that is sufficient to generate all differential invariants by means of acting on the generating invariants with the operators of invariant differentiation and taking combinations of the basis invariants with these invariant derivatives. Often the computation of the \emph{syzygies} among the differential invariants is a crucial step to prove the minimality of a given generating set. The two operators of invariantization for the maximal Lie invariance group~$G$ of the KdV equation follow from the invariantization of the operators of total differentiation $\mathrm D_t$ and $\mathrm D_x$ and they are
\begin{gather*}
 \mathrm D_t^{\rm i}=\iota(\mathrm D_t)=(u_t+uu_x)^{-3/5}(\mathrm D_t+u\mathrm D_x),\\
 \mathrm D_x^{\rm i}=\iota(\mathrm D_x)=(u_t+uu_x)^{-1/5}\mathrm D_x.
\end{gather*}

In~\cite{CardosoBihlo:cheh08Ay} it was claimed that the invariants
\[
 I_{01}=\frac{u_x}{(u_t+uu_x)^{3/5}},\quad I_{20}=\frac{u_{tt}+2uu_{tx}+u^2u_{xx}}{(u_t+uu_x)^{8/5}}
\]
form a generating set of the algebra of differential invariants for the KdV equation. While this is certainly true, this set is not minimal. In~\cite{CardosoBihlo:olve10Ay} it was shown that the differential invariant $I_{01}$ is in fact sufficient to generate the entire algebra of differential invariants for the KdV equation. The crucial step missed in finding the minimal generating set in~\cite{CardosoBihlo:cheh08Ay} was the use of the commutator formula for the operators of invariant differentiation $\mathrm D_t^{\rm i}$ and $\mathrm D_x^{\rm i}$, which is
\begin{equation}\label{CardosoBihlo:CommutationRelationKdV}
 [\mathrm D_t^{\rm i},\mathrm D_x^{\rm i}]=\tfrac35(I_{11}+I_{01}^2)\mathrm D_t^{\rm i}-\tfrac15(I_{20}+6I_{01})\mathrm D_x^{\rm i}.
\end{equation}
From the recurrence relation
\[
 \mathrm D_t^{\rm i}I_{01}=-\tfrac35I_{01}^2+I_{11}-\tfrac35I_{01}I_{20}
\]
one can solve for $I_{11}$ in terms of $I_{01}$ and $I_{20}$. Applying the commutation relation~\eqref{CardosoBihlo:CommutationRelationKdV} to the invariant $I_{01}$ then allows solving for $I_{20}$ solely in terms of $I_{01}$, which explicitly gives
\[
 I_{20}=\frac{[\mathrm D_t^{\rm i},\mathrm D_x^{\rm i}]I_{01}-\frac35(\mathrm D_t^{\rm i}I_{01}+\frac85I_{01}^2)\mathrm D_t^{\rm i}I_{01}+\frac65I_{01}\mathrm D_x^{\rm i}I_{01}}{\frac9{25}I_{01}\mathrm D_t^{\rm i}I_{01}-\frac45\mathrm D_x^{\rm i}I_{01}},
\]
which shows that $I_{01}$ is indeed the minimal generating set of the algebra of differential invariants for the KdV equation.

We now repeat the computation of a basis of differential invariants for the moving frame~\eqref{CardosoBihlo:MovingFrameKdVequation2}. The associated operators of invariant differentiation for this moving frame are the same that were constructed in~\cite{CardosoBihlo:chup04Ay} within the framework of the infinitesimal approach,
\[
 \mathrm D_t^{\rm i}=u_x^{-1}(\mathrm D_t+u\mathrm D_x),\quad \mathrm D_x^{\rm i}=u_x^{-1/3}\mathrm D_x.
\]

The computation of corresponding recurrence relations differs from that given in~\cite{CardosoBihlo:cheh08Ay,CardosoBihlo:olve10Ay}
only in minor details. Identifying $c_3=\xi_t$ and $c_4=\frac13\tau_t$, we obtain the invariantized forms
\begin{gather*}
\hat\tau=\iota(\tau),\quad
\hat\xi=\iota(\xi),\quad
\hat\eta=\hat\eta^{00}=\iota(\eta),\\
\hat\eta^\alpha=\iota(\eta^\alpha)=-\frac{3\alpha_1+\alpha_2+2}3I_\alpha\hat\tau^1-\alpha_1I_{\alpha_1-1,\alpha_2+1}\hat\eta, \quad \alpha_1+\alpha_2>0,
\end{gather*}
and the first three forms~$\hat\tau$, $\hat\xi$ and $\hat\eta$ jointly with $\hat\tau^1=\iota(\tau^1)$ make up,
in view of the invariantized counterpart of the determining equations~\eqref{CardosoBihlo:DetEqs},
a basis of the invariantized Maurer--Cartan forms of the algebra~$\mathfrak g$.
The recurrence formulas for the normalized differential invariants are
\[
\mathrm d_{\mathrm h}H^1=\omega^1+\hat\tau, \quad
\mathrm d_{\mathrm h}H^2=\omega^2+\hat\xi, \quad
\mathrm d_{\mathrm h}I_{\alpha}=I_{\alpha_1+1,\alpha_2}\omega^1+I_{\alpha_1,\alpha_2+1}\omega^2+\hat\eta^\alpha,
\]
where the form $\omega^1=\iota(\mathrm dx)$ and $\omega^2=\iota(\mathrm dx)$ constitute the associated invariantized horizontal co-frame,
$\mathrm d_{\mathrm h}$ is the horizontal differential and so
$\mathrm d_{\mathrm h}F=(\mathrm D_t^{\rm i}F)\omega^1+(\mathrm D_x^{\rm i}F)\omega^2$.
We take into account that $H^1=0$, $H^2=0$, $I_{00}=0$ and $I_{01}=1$ and solve the corresponding recurrence formulas with respect to
the basis invariantized Maurer--Cartan forms,
\[
\hat\tau=-\omega^1, \quad
\hat\xi=-\omega^2, \quad
\hat\eta=-I_{10}\omega^1-\omega^2,\quad
\hat\tau^1=I_{11}\omega^1+I_{02}\omega^2.
\]
Then splitting of the other recurrence formulas yields
\begin{gather*}
\mathrm D_t^{\rm i}I_\alpha=I_{\alpha_1+1,\alpha_2}-\frac{3\alpha_1+\alpha_2+2}3I_{11}I_\alpha+\alpha_1I_{10}I_{\alpha_1-1,\alpha_2+1},\\
\mathrm D_x^{\rm i}I_\alpha=I_{\alpha_1,\alpha_2+1}-\frac{3\alpha_1+\alpha_2+2}3I_{02}I_\alpha+\alpha_1I_{\alpha_1-1,\alpha_2+1},
\end{gather*}
where $\alpha_1>0$ or $\alpha_2>1$.

It is obvious from the above split recurrence formulas for that the whole set of differential invariants of the maximal Lie symmetry group of the KdV equation is generated by the two lowest-order normalized invariants
\[
I_{10}=u_x^{-5/3}(u_t+uu_x),\quad I_{02}=u_x^{-4/3}u_{xx}.
\]
At the same time, the differential invariant~$I_{02}$ is expressed in terms of invariant derivatives of~$I_{10}$
and hence a basis associated with the moving frame~\eqref{CardosoBihlo:MovingFrameKdVequation2} consists of the single element~$I_{10}$.
Indeed, we have
\[
[\mathrm D_x^{\rm i},\mathrm D_t^{\rm i}]=-I_{02}\mathrm D_t^{\rm i}+(1+\tfrac13I_{11})\mathrm D_x^{\rm i}
=-I_{02}\mathrm D_t^{\rm i}+(\tfrac13(\mathrm D_x^{\rm i}I_{10})+\tfrac59I_{10}I_{02}+\tfrac23)\mathrm D_x^{\rm i}
\]
as $I_{11}=\mathrm D_x^{\rm i}I_{10}+\tfrac53I_{10}I_{02}-1$.
Applying the commutation relation for $\mathrm D_x^{\rm i}$ and $\mathrm D_t^{\rm i}$ to~$I_{10}$
and solving the obtained equation with respect to~$I_{20}$, we derive the requested expression,
\[
I_{20}=\frac
{[\mathrm D_t^{\rm i},\mathrm D_x^{\rm i}]I_{10}-\frac13(\mathrm D_x^{\rm i}I_{10}+2)\mathrm D_x^{\rm i}I_{10}}
{\frac59I_{10}\mathrm D_x^{\rm i}I_{10}-\mathrm D_t^{\rm i}I_{10}}.
\]

\section{Conclusion}\label{CardosoBihlo:ConclusionKdV}

The present paper is devoted to the construction of differential invariants for the maximal Lie invariance group of the KdV equation. We illustrate by examples that it is worthwhile to examine different possibilities for choosing the normalization conditions, which is a cornerstone for the moving frame computation. This is an important investigation as the form of differential invariants obtained depends strongly on the set of normalization equations chosen. In the present case of the maximal Lie invariance group of the KdV equation, using $U_X=1$ as a~normalization condition instead of the condition $U_T=1$ chosen in~\cite{CardosoBihlo:cheh08Ay} leads to the normalized differential invariants~\eqref{CardosoBihlo:DifferentialInvariantsKdVSecondNormalization} which have a simpler form than the normalized differential invariants~\eqref{CardosoBihlo:DifferentialInvariantsKdVFirstNormalization} associated with the latter condition.
The same claim is true concerning the corresponding operators of invariant differentiation, recurrence formulas, etc.
Moreover, the differential invariants~\eqref{CardosoBihlo:DifferentialInvariantsKdVSecondNormalization} are singular only on a~proper subset of the set of solutions of the KdV equation for which the differential invariants~\eqref{CardosoBihlo:DifferentialInvariantsKdVFirstNormalization} are singular.
The invariantized form of the KdV equation is more appropriate using the normalization condition $U_X=1$.
In contrast to the condition~$U_T=1$, this condition also naturally leads to the separation of differential invariants which involve only derivatives of~$u$ with respect to~$x$ that may be essential as the KdV equation is an evolution equation.

We also show that for Lie groups of rather simple structure, it is possible to construct functional bases of differential invariants of arbitrary order in an explicit and closed form like~\eqref{CardosoBihlo:DifferentialInvariantsKdVFirstNormalization} and~\eqref{CardosoBihlo:DifferentialInvariantsKdVSecondNormalization}. This observation was first presented in~\cite{CardosoBihlo:bihl11Fy} for an infinite-dimensional Lie pseudogroup.
Such a closed-form expression is beneficial as it is generally simpler than the form of differential invariants obtained when acting with operators of invariant differentiation on basis differential invariants. It is difficult to conceive finding similar expressions for arbitrary order within the framework of the infinitesimal method in a reasonable way.

\section*{Acknowledgements}

The author is grateful to Alexander Bihlo and Roman O.\ Popovych for helpful discussions and suggestions.
This research was supported by the Austrian Science Fund (FWF), project P20632.

\LastPageEnding

\end{document}